\documentclass[12pt,preprint]{aastex}
\input{epsf}
\begin{document}

\newcommand{\lap}{$L_{38}^{-1/3}$}
\newcommand{\ergs}{\rm \su  erg \su s^{-1}}
\newcommand{\etal}{ {\it et al.}}
\newcommand{\porb}{ P_{orb} }
\newcommand{\Po}{$P_{orb} \su$}
\newcommand{\pdot}{$ \dot{P}_{orb} \,$}
\newcommand{\pot}{$ \dot{P}_{orb} / P_{orb} \su $}
\newcommand{\mm}{$ \dot{m}$ }
\newcommand{\mdot}{$ |\dot{m}|_{rad}$ }
\newcommand{\myr}{ \su M_{\odot} \su \rm yr^{-1}}
\newcommand{\msol}{\, M_{\odot}}
\newcommand{\ppp}{ \dot{P}_{-20} }
\newcommand{\cms}{ \rm \, cm^{-2} \, s^{-1} }
\newcommand{\pdott}{ \left( \frac{ \dot{P}/\dot{P}_o}{P_{1.6}^{3}}\right)}
\newcommand{\be}{\begin{equation}}
\newcommand{\ee}{\end{equation}}
\newcommand{\nn}{\mbox{} \nonumber \\ \mbox{} }
\newcommand{\ba}{\begin{eqnarray}}
\newcommand{\ea}{\end{eqnarray}}

\def\p{\phantom{1}}
\def\pmu{\mox{$^{-1}$}}
\def\kms{km^s$^{-1}$}
\def\sbu{mag^arcsec${{-2}$}}
\def\e{\mbox{e}}
\def\dex{\mbox{dex}}
\def\L{\mbox{${\cal L}$}}
\def\gte{\lower 0.5ex\hbox{${}\buildrel>\over\sim{}$}}
\def\lte{\lower 0.5ex\hbox{${}\buildrel<\over\sim{}$}}
\def\loe{\lower 0.6ex\hbox{${}\stackrel{<}{\sim}{}$}}
\def\goe{\lower 0.6ex\hbox{${}\stackrel{>}{\sim}{}$}}

\shorttitle{Magnetars: Evolution and Field Decay}
\shortauthors{Arras, Cumming, and Thompson}
\title{
Magnetars: Time Evolution, Superfluid Properties, and Mechanism
of Magnetic Field Decay}

\author{
P. Arras\altaffilmark{1}, 
A. Cumming\altaffilmark{2},
C. Thompson\altaffilmark{3}
}
\altaffiltext{1}{Kavli Institute for Theoretical Physics,
University of California, Santa Barbara, CA}
\altaffiltext{2}{Department of Astronomy and Astrophysics,
University of California, Santa Cruz, CA}
\altaffiltext{3}{Canadian Institute for Theoretical Astrophysics,
60 St. George St., Toronto, ON M5S 3H8}

\begin{abstract}
We calculate the coupled thermal evolution and
magnetic field decay in relativistic model neutron stars threaded
by superstrong magnetic fields ($B > 10^{15}$ G).  Our main goal
is to evaluate how such ``magnetars'' evolve with time and
how field decay modifies the transitions to core
superfluidity and cooling dominated by surface X-ray emission.
Observations of a thermal X-ray spectral component and fast
timing noise place strong constraints on the presence of a superfluid core.
We find that the transition to core superfluidity can be significantly
delayed by field decay in the age range $\sim 10^3-10^5$ yrs.  The mechanism
of Hall drift is related to the stability of the core magnetic field,
and to currents flowing outward through the crust.
The heating effect is enhanced if it is continuous rather than
spasmodic.  Condensation of a heavy element layer at the surface is shown
to cause only modest changes in the outward conduction of heat.

\end{abstract}

\section{Introduction}

Observations of the Soft Gamma Repeaters and Anomalous X-ray Pulsars 
have revealed a rich phenomenology of transient X-ray emission and
torque variability.  Some members of the first group
occasionally emit enormously 
bright X-ray flares (Hurley 2000), which are followed by transient
periods of decaying X-ray flux.  The unified nature of the SGRs and
AXPs is indicated by the detection of $\sim 0.1$ s hard-spectrum
X-ray bursts from two AXPs (Gavriil et al. 2002; Kaspi et al. 2003).  
This variable X-ray emission is now generally believed to be powered
by the decay of an ultrastrong magnetic field
(Thompson \& Duncan 1996; Colpi, Geppert \& Page 2000).

Two sub-classes of these objects have
emerged recently:  those which maintain a fairly steady X-ray luminosity
over two decades or longer; and a less conspicuous group (not 
necessarily smaller in total numbers) in which one sees
transitions to and from very low levels of persistent X-ray emission
over a period of months to years (Torii et al. 1998; Kouveliotou 
et al. 2003; Ibrahim et al. 2003).  Objects in the second group are 
weak burst sources or have not been observed to burst at all.

Some SGRs and AXPs also show strong timing irregularities,
which broadly can be divided into i) 
adiabatic variations in torque over a period
of months to years (Kaspi et al. 2001; Woods et al. 2002) and
ii) a broad spectrum of timing noise starting from periods as short 
as $\sim 10^4$ s and upward.  Our goal here is to isolate the most 
important mechanisms by which an ultrastrong magnetic field
will contribute to the persistent X-ray emission of these objects
(on timescales of years or longer);  and also to make deductions 
about the composition of the star from the appearance 
of strong timing noise in active burst sources.

\section{Basic Modes of Magnetic Field Decay}

A decaying magnetic field in a neutron star evolves through
a series of equilibrium states, punctuated by the release
of elastic stresses in its crust and the excitation of
hydrodynamic motions in its liquid core.  The crust is too weak to sustain
all but modest departures from magnetostatic equilibrium, if
$B > 10^{15}$ G.   In the case of
a purely fluid star that is stabilized against convection in 
the radial direction, it is generally believed that the equilibrium
states of the magnetic field must carry some net helicity, e.g., 
that the field has both toroidal and poloidal components.  Purely 
poloidal (Flowers and Ruderman 1975) and purely toroidal 
(Tayler 1973) fields are unstable.   

The magnetic field evolves according to the equation
\be\label{induction}
{\partial{\bf B}\over \partial t} = {\bf\nabla}\times
\left[\left({\bf v}+{\bf v}_{\rm amb}+{\bf v}_{\rm Hall}\right)
\times{\bf B}\right] + 
{\partial{\bf B}\over \partial t}\biggr|_{\rm fracture}
\ee
Here ${\bf v}_{\rm amb}$ is the speed with which the magnetic field
is advected by the diffusing charged component (electrons/protons) of
the neutron star core; ${\bf v}_{\rm Hall} = -{\bf J}/en_e$; and
${\bf v}$ is the hydrodynamic response of the core to the combined
effect of these transport processes.  The rigid crust is also 
subject to sporadic yields and fractures which cause changes in
${\bf B}$ (both in crust and core) on short timescales.
The star is spherical in a first approximation, and the compositional
stratification enforces 
$v_r = O(\varepsilon_B) \simeq 0$ and
${\bf\nabla}\cdot{\bf v} = O(\varepsilon_B) \simeq 0$
(Reisenegger \& Goldreich 1992).
Here $\varepsilon_B = B^2/8\pi P_e$ and $P_e$ is the electron pressure.

The diffusion of charged particles in a normal, degenerate $n$-$p$-$e$ plasma
is limited by proton-neutron drag at high temperatures (Haensel et al.
1990; Goldreich \& Reisenegger 1992, hereafter GR); 
and at $T < 5\times 10^8$ K by the
rate of relaxation of chemical potential gradients
(GR; Pethick 1992).   
We note here that the $n$-$p$ collision
rate is also reduced dramatically after the transition to proton
superconductivity (due to the lower density of quasiparticle
excitations near the proton fermi surface).
This probably happens early on compared with the
$\sim 10^4$ yr lifetime of SGR/AXP activity
(e.g. Yakovlev et al. 2001).  To calculate the continuing ambipolar
diffusion of a magnetic field in a superconducting core, we therefore 
include only the limiting effects of chemical potential gradients.
In this case, the timescale is 
$t_{\rm amb}(B,T) \simeq$ ${8\pi n_e^2 (k_BT)^2/\dot U_{\rm Urca}(\rho,T) B^2}$
when the chemical potential imbalance $\Delta \mu \sim B^2/8\pi n_e$
induced by the ${\bf J}\times{\bf B}/c$ force is much less than $k_B T$.
Here $\dot U_{\rm Urca}$ is the rate at which beta transformations
between neutrons and protons release energy to neutrinos.  This
rate is strongly modified by Cooper pairing of protons and neutrons,
and we include the resulting corrections as tabulated in
Yakovlev et al. (2001).  For mean fields stronger than several 
$\times 10^{14}$ G, the magnetic sheaths of the superconducting fluxoids
are packed together.  Any collective motion of the fluxoids with respect
to the protons is strongly inhibited, and we can treat the field as being
continuous.  In a normal $n$-$p$-$e$ core,
the heating induced by the decay of the field leads to a direction 
relation $k_BT \sim \Delta\mu \propto B^2$ between $T$ and $B$.  Given
the strong $T$-dependence of the URCA rates, this causes a strong feedback 
$t_{\rm amb} \propto B^{-14}$ on the drift rate above a critical 
flux density of $\sim 3\times 10^{15}$ G (Thompson \& Duncan 1996).

If the interior magnetic field of a neutron star is helical, then
the current density ${\bf J}$ and ${\bf B}$ both have poloidal components.
In a magnetar, closure of this poloidal current inside the star
generates a ${\bf J}\times{\bf B}/c$ force which is strong enough to 
fracture the crust, thereby twisting up the external magnetic field
(Thompson, Lyutikov, \& Kulkarni 2002).  This mechanism is driven 
by Hall effect in the crust, through the
gradient in electron density with height $z$.  The Hall term in eq.
(\ref{induction}) yields
$t_{\rm Hall}^{-1} \equiv$ $B_\phi^{-1}(\partial B_\phi/\partial t) = 
J_z (\partial/\partial z)(en_e)^{-1}$
in cylindrical coordinates.
Where the poloidal current flows outward -- $J_z > 0$ --
the toroidal field will migrate toward lower 
densities.  There the crust has a smaller shear modulus and is less 
able to balance magnetic shear stresses.  The net effect is to excavate
the twist in the poloidal field from the crust, on the timescale
$t_{\rm Hall} =  2.4\times 10^5\,(R_6/B_{\phi,15}) \rho_{14}^{5/3}$ yr
at a radius $R = R_6\times 10^6$ cm and density
$\rho_{14}\times 10^{14}$ g cm$^{-3}$.  

The Hall effect has different consequences in the fluid core.
In a liquid that is stratified parallel to gravity ${\bf g}$, 
Hall drift creates unbalanced magnetic stresses.  In particular,
hall waves do not exist as propagating modes in the high frequency limit, 
$k|d\ln\rho/dz|^{-1} \gg 1$.  To show this, note that the Hall term
can only be cancelled by a hydrodynamic displacement satisfying $v_z = 0$.
This condition is satisfied if ${\bf v}_{\rm Hall}$ has no vertical 
component, i.e., if$J_z = 0$. 
A simple example is a finite-amplitude field variation
${\bf B}^\prime = {\bf B}^\prime_0 \exp(i{\bf k}\cdot{\bf x})$ 
superimposed on a uniform background field ${\bf B}_0$.
Magnetostatic equilibrium requires that ${\bf B}^\prime$ be linearly
polarized, and that $({\bf B}_0^\prime\times{\bf k})\cdot{\bf g} =0$,
which implies in turn $J_z =0$.

On larger scales, the Hall term in the
induction equation generally {\it cannot} be cancelled off by such a
hydrodynamic displacement field in the fluid 
core.\footnote{Mestel (1956) has argued that Hall drift of an 
axisymmetric poloidal magnetic field will excite a torsional Alfv\'en 
wave in a fluid star.  However, such a field configuration has a 
special symmetry (radial current $J_r = 0$) and the above argument 
indicates that the Hall drift will instead be cancelled by a compensating
hydrodynamic motion.}
Hall drift evolves the magnetic field into a new configuration of
equal total energy (GR),
but if the field is initially an a stable 
configuration then all displacements satisfying 
$v_r = 0$, ${\bf\nabla}\cdot{\bf v} = 0$
increase its energy.  
In fact, the new configuration
will generally have a {\it higher} energy than some neighboring 
configuration into which it can relax. In what follows, 
we make the reasonable assumption that this excess energy
is converted to heat on a timescale much less than $\sim 10^3$ yrs.

The next
step is a prescription for the {\it hydrodynamic} response of
the fluid interior to this diffusive motion.  
We define a characteristic poloidal magnetic field 
$B_P$, toroidal field $B_\phi = O(10)B_P$, and tilt angle
$\alpha$ between the poloidal current and the direction of
gravity at the center of each toroidal loop.  There is a characteristic
tilt, $\alpha \sim B_P/B_\phi$,
below which the field is able to largely unwind through 
differential rotations of fluid shells confined to 
gravitational equipotential surfaces (Fig. 2 of Thompson \& Duncan 2001).
Hence changes in $\alpha$ can cause unwinding of the toroidal
field at the rate $dB_\phi/dt = -(B_\phi/\alpha)\,|d\alpha/dt|$.
On the other hand, microscopic transport of the field on a
timescale $\tau$ causes changes in $\alpha$ at a rate $d\alpha/dt \sim \pm
\tau^{-1}$.  Thus we evolve the simple one-field model
\be\label{bphieq}
{dB_\phi\over dt} = -{B_\phi\over\alpha}\left[{1\over t_{\rm amb}(B_\phi)} +
{1\over t_{\rm Hall}(B_\phi)}\right].
\ee
Unwinding of the field does not require any change in $B_P$, and the
associated transport time is much longer when $B_P \ll B_\phi$. 
Hence we only evolve $B_\phi$.

A net change in winding of the core magnetic field must be accompanied
by a torsional deformation of the stellar crust.  The energy deposited
in an area $A$ of the crust by twisting it through an angle $\theta$ is 
$\sim \theta^2 \int \mu  dV \sim 5\times 10^{40}\,(\theta/0.001)^2\,
(A/100~{\rm km}^2)$ ergs.  (Here $\mu = 1.1\times 10^{30}\rho_{14}^{0.8}$ is
the crustal shear modulus; Strohmayer et al. 1991.)  During a 
concentrated episode of SGR activity, when hundreds of 
$\sim 0.1$-s X-ray bursts are emitted, this process might 
be repeated $\sim \theta^{-1}$ times, and the net crustal 
heat deposition would be $\sim 1000\,(\theta/0.001)^{-1}$ times larger.  
A giant flare evidently involves a single readjustment through
a much larger angle.

A toroidal magnetic field confined to the crust will,
in a first approximation, evolve independently from that in the core.
The net angle through which a radial poloidal field is twisted
(across one vertical density scale height) is limited by the finite
yield strain $\psi$ of the crust,
$\Delta\phi_B = (\ell_\rho/R)(B_\phi/B_z) \leq (4\pi\mu/B_z^2)\psi$.
This sets an upper bound $B_\phi
\leq 7\times 10^{15}\,B_{z,14}^{-1}(\psi/10^{-3})R_6$ G.
Integrating the above Hall equation for $B_\phi$, under the assumption
that inhomogeneities in the winding of the field relax completely in
the deep crust after each yielding event, gives
$B_\phi/B_{\phi,0} = 
[1+ t/t_{\rm Hall}(B_{\phi,0})]^{-1}$.
The corresponding heating rate, integrated over depth $z$,
$2\times 10^{33}
\,B_{\phi,15}^3A_{13} R_6^{-1}
1+t/t_{\rm Hall}(B_{\phi,0})]^{-3}\;{\rm ergs~s^{-1}}$,
is constant as $t\rightarrow 0$ and scales as $t^{-3}$ at 
$t \gg t_{\rm Hall}(B_{\phi,0})$.   We will revisit the interplay
between radial currents and Hall drift elsewhere in more detail.

\section{Coupled Thermal Evolution and Field Decay}

The thermal evolution of the star is followed using a relativistic model
constructed from the Tolman-Oppenheimer-Volkoff 
equations and the parametrized equation of
state of Prakash, Lattimer, and Ainsworth (1988)
with intermediate compressibility ($K = 240$ MeV)
and a mass low enough that direct URCA cooling can be neglected ($M = 1.35
\, M_\odot$).
(We note that more recent EOS indicate the absense of direct URCA cooling
for much larger masses;  Akmal, Pandharipande, and Ravenhall 1998.)
We are interested here in baseline thermal evolution, averaged over 
a timescale long compared with the conduction times across the crust,
which is more than two orders of magnitude shorter than the spindown
ages of the SGRs and AXPs ($P/2\dot P \ga 10^3$ yr).  
The calculation therefore takes into account the effects of continuous
modes of magnetic field decay.

We calculate the net neutrino luminosity $L_\nu$ by integrating the 
emissivities due to the
modified URCA reaction ($\{n,p\}+n
\leftrightarrow \{n,p\} + p + e^- + \bar\nu_e$), nucleon bremsstrahling 
emission ($n + n \rightarrow n + n +\nu + \bar\nu$, etc.) and neutron 
Cooper pair emission ($n + n \rightarrow (2n) + \nu + \bar\nu$);
evolving the magnetic field according to eq. (\ref{bphieq}) using
the volume-averaged modified-URCA rate in $t_{\rm amb}$;
calculating the luminosity $L_X = 4\pi R_{\rm NS}^2 e^{2\phi(R_{\rm NS})}
\sigma_{\rm SB} T_s^4$ ($\phi(r) =$ gravitational potential)
in thermal surface X-ray emission from
the relation between core temperature and surface temperature $T_s$ detailed
below; and, finally, evolving the (redshifted) core temperature 
$T_c = e^{\phi(r)} T(r)$ in the isothermal approximation by
balancing the net cooling luminosity with the rate
of loss of magnetic energy from the volume $V$ of the star,
\be\label{tempev}
\langle C_V\rangle{dT_c\over dt} 
= -{1\over V}\Bigl(L_\nu + L_X\Bigr) + 
e^{2\phi(R_{\rm NS})}{B_\phi\over 4\pi}{dB_\phi\over dt}.
\ee
Balancing only the last term with the 
modified-URCA emissivity, one finds $\Delta \mu \sim B^2/8\pi n_e
\sim \alpha (k_BT_c)$.   There are finite-$\Delta\mu$ 
corrections to the URCA rates (Reisenegger 1995), but these are
generally less important than the corrections due to nucleon pairing
when $\alpha \ll 1$.

The heat flux emerging through the surface of a magnetar depends on
the relation between the surface effective temperature
$T_s$  and the temperature $T_c$
in the deep crust.  This relation has been calculated (Van Riper 1988;
Heyl \& Hernquist 1997; Potekhin and Yakovlev 2001) for a magnetized
atmosphere in which the ions form a non-ideal gas.  At the
surface temperature characteristic of SGRs and AXPs ($k_BT_{\rm bb} 
\sim 0.4$-$0.5$ keV; \"Ozel, Psaltis, \& Kaspi 2001) 
one however expects heavy ions (e.g. iron) to be 
condensed into long molecular chains (Lai and Salpeter 1997).  The 
density is then large even close to the surface
and can be estimated by minimizing the sum of the Coulomb 
and electron degeneracy energies in a Wigner-Seitz 
cell:\footnote{Here $Y_e$ 
is the electron fraction
and $Z$ the nuclear charge.} 
$\rho(0) \simeq 1.8\times 10^7 (Y_e/0.5)^{-1} (Z/26)^{2/5} 
(B/10^{15}~{\rm G})^{6/5}$ g cm$^{-3}$. 

The transmission of heat through the atmosphere of a neutron star
with zero surface magnetic field is controlled by a sensitivity strip
of a lower density $\rho \sim 10^5-10^6$ g cm$^{-3}$ (e.g. Potekhin \&
Yakovlev 2001).  Fig. 1 shows the surface temperature of a neutron star
with normal core neutrons and protons (transition temperatures 
$T_{c,n} = T_{c,p} = 0$) using this standard envelope
model.  We have also included the effect of ambipolar diffusion of
an internal toroidal field ($B_\phi = 3$, $5\times 10^{15}$ G),
using eq. (2) and the ambipolar diffusion timescale $t_{\rm amb}$
tabulated in eqs. (58), (59) of GR. (See also Heyl \& Kulkarni
1998.)

The heat flux is higher in a second envelope
model which we have constructed for a condensed
iron surface layer in a $10^{15}$ G magnetic field.  The best power-law fit
is $T_s/g_{14}^{1/4} 
= 4.0\times 10^6~{\rm K}\,({T_c/10^9~{\rm K}})^{1/2}$
[$2.9\times 10^6~{\rm K}\,({T_c/10^9~{\rm K}})^{1/4}$]
for $T_c$ greater than [less than] $2.8\times 10^8 $K, and 
gravity $g = g_{14}\times 10^{14}$
cm s$^{-2}$.  The integration of $dT/dz$ over depth $z$ idealizes the
electrons as a (locally) uniform fermi gas, and employs the
thermal conductivity tabulated by Potekhin (1999). The ideal
finite-$T$ electron equation of state is supplemented by
a correction for the (negative) Coulomb pressure in the solid
phase, $P = P_e - P_e(0)[n_e/n_e(0)]^{4/3}$, where $P_e(0)$
and $n_e(0)$ are the electron degeneracy pressure and density
at zero total pressure $P$.  

Much of the thermal resistance is localized where the electrons
are 1-dimensional, with fermi momentum $p_{Fe} \sim m_ec$ 
(density $\rho = Y_e m_p eBp_{Fe}/2\pi^2\hbar^2c \sim 
10^8B_{15}$ g cm$^{-3}$) and where the
melting temperature is $1\times 10^8
(Z/26)^{5/3}B_{15}^{1/3}$ K.  Thus at high $T_c$
the sensitivity strip lies in a Coulomb liquid layer below
the solid surface; whereas at low $T_c$ (and large
$Z$) it lies within the solid.   In the solid, the electrons
are nearly degenerate and heat is transported by conduction
electrons within a narrow energy range $\sim k_BT$ near the fermi surface.  
The scalings for $T_s(T_c)$
are easily derived by approximating the 
conductivity as being due to phonon scattering (in the solid)
or Coulomb scattering (in the liquid), with all the electrons
in the lowest Landau state.
This analytic model gives $T_s^4 \propto B_s^{1/5}Z^{-3/5}$ at high
$T_c$, and $T_s^4 \propto B_s$ at low $T_c$.  To correct for
the lower opacity of the extraordinary mode near the condensed
surface, we show $2^{1/4}$ times the effective temperature.
Since there is evidence for complicated multipolar structure in 
the surface magnetic field of some SGRs (Feroci et al. 2001), $B_s$
is a characteristic surface field without any dipole structure.

Our numerical results are plotted in Figs. 1 and 2. In sum:

i) Steady heating of the stellar interior by a $\sim 3\times 10^{15}$ G
magnetic field increases $\log_{10}(T_s)$ by $\sim 0.2$ if the core
neutrons and protons are normal.  Proton pairing suppresses the
rate of charged-current weak interactions and therefore the rate
of the irrotational mode of ambipolar diffusion; in which case Hall drift
makes a similar contribution to the decay of the core magnetic field before
the core superfluid transition, and dominates thereafter.

ii) If the peak pairing temperature $T_{c,n} \ga 6\times 10^8$ K,
the drop in $T_s$ is gradual and occurs earlier than $\sim 100$ years;
but if $T_{c,n} \la 5\times 10^8$ K, then this drop is sharper and
occurs in the observed age range of SGR/AXP activity.  Magnetic dissipation
can delay the time of the pairing transition by an order of magnitude.
It also delays (and makes sharper) the transition to photon-dominated cooling.
If the internal B-field is much stronger than $10^{14}$ G,
{\it and} its coherence length is large ($\sim 10$ km), 
then continuous Hall decay 
(which is not temperature sensitive) will keep the surface warm 
up to an age of several $\times 10^5$ yrs (redshifted
temperature $T_s^\infty = T_s e^{\phi(R_{\rm NS})} \sim 1.5\times 10^6$ K).
Strong intermittency in the rate of
Hall decay would allow the core to cool off more rapidly in between
magnetic `flares', due to the strong temperature dependence of the
neutrino emissitivities, and the temperature evolution would be intermediate
between the dashed and solid curves in Fig. 2.

iii) The thermal evolution with a condensed Fe atmosphere gives
 $T_s^\infty$ within a factor $\sim 1.3$-$1.5$ of those measured in the thermal
components of AXP spectra ($kT_{\rm bb} \simeq 0.4$-$0.5$ keV) at an
age of $\sim 10^3$-$10^4$ yrs.  A relatively thin light-element
layer will harden the thermal peak by this amount, compared with a pure
black body (e.g. Lloyd, Hernquist, and Heyl 2003).  If the surface
field is $\sim 10^{15}$ G, then $L_X$ is generally less than 
$10^{35}$ erg s$^{-1}$, consistent with some AXP sources but not all
(e.g. \"Ozel et al. 2001).  Recalculating the cooling models with
with a lighter element (carbon) surface layer
shows a factor $\sim 3-4$ increase in thermal transparency at early times
-- but not below $T_s \sim 4\times 10^6$ K, so that the 
photon cooling time does not change significantly.
Heating by an external
current may also not be negligible:  the power needed to 
force a current of ions and electrons through a twisted magnetosphere 
with a constant equatorial pitch $B_\phi/B_\theta$ is
$L_X \sim 3\times 10^{35}$ $(B_{\rm pole}/10^{14}~{\rm G})
(B_\phi/B_\theta)$ ergs/s (Thompson et al. 2002).

\section{Timing Noise and Core Neutron Superfluidity}

Phase-resolved X-ray timing of SGRs 1806$-$20 and 1900$+$14 have
revealed a broad spectrum of timing noise (Woods et al. 2002).
At short time invervals ($\Delta t\sim 3\times 10^4$ s) large phase offsets
($\Delta\phi \sim 0.3$ cycles) are sometimes observed in the X-ray
pulses, which prevent phase-connected timing.  These offsets correspond
to stochastic shifts in frequency of magnitude 
$\Delta\nu/\nu \sim 10^{-4}$.  The alternative is an enormous 
increase in frequency derivative by $\sim 10^2$ ($\Delta\dot\nu
\sim 2\Delta\phi/\Delta t^2 \sim 10^{-9}$ s$^{-2}$).  
Hence the effect appears to be due to the absorption
and release of angular momentum by an internal superfluid component.
The lack of obvious classical post-glitch relaxation behavior, and 
the magnitude of the timing residuals, lead us identify a {\it superfluid core}
as the reservoir of angular momentum.  Indeed, no obvious correlation 
exists between this timing noise and the observation of long-term torque
variations, bright X-ray outbursts, or large changes in persistent 
X-ray flux that would be associated with deformations of the crust.  

Timing noise of this amplitude is absent in
radio pulsars, and its presence in magnetars is strongly
correlated with {\it overall} activity as a burst source, and with
a spectrally hard persistent X-ray emission.   Part of this
correlation could result from the reduction in interior temperature
(by a factor $\sim 5$) coinciding with the pairing transition of
the neutrons, which would make the crust more brittle; and from
a reduction in surface cooling rate in comparison with the power
dissipated by external currents.  

In a slowly rotating magnetar, the outward motion of the vortices 
is constrained by the magnetic field, due to  the large 
energetic barrier to the crossing of vortices and superconducting 
fluxoids (e.g. Ruderman et al. 1998).   Slow
fluctuations in the field of amplitude 
$\Delta B/B \sim (I_{\rm sf}/I)^{-1}\Delta\nu/\nu \sim 10^{-4}
(I_{\rm sf}/I)^{-1}$, with some amplitude perpendicular to the axis
of rotation, would provide the necessary perturbation to the superfluid.
If such fluctuations are to occur over many ($> 10^4$)
core Alfv\'en-crossing times, the field configuration 
must be metastable, and nearly degenerate in energy with others
that differ by one part in $\sim 10^4$.  This is expected if
the core field has relaxed significantly from
an initial equilibrium state.

\section{Conclusions and Observational Tests}

Several degrees of freedom strongly influence the thermal and
spectral evolution of a magnetar:   i) the strength of its
internal magnetic field; ii) the temperature at which core 
neutrons become superfluid; iii) the transmissivity of its
thermal envelope; and iv) the configuration of its internal magnetic
field and the equilibrium states through which this field moves.
Before a core superfluid transition, thermal luminosities up to
$\sim 10^{35}$ ergs/s are possible if the surface magnetic field
is $\sim 10^{15}$ G; thereafter, the neutrino emissivity rises
dramatically due to Cooper pair cooling, and thermal emission powered
by internal Hall decay can continue at a level $\la 10^{34}$ ergs/s
beyond an age of $\sim 10^5$ yrs.
We infer that magnetar candidates which show transitions to/from
such low luminosity states have magnetic fields of intermediate strength
and/or superfluid cores.  We have also argued that
the observation of fast timing noise in SGRs 1900$+$14 and 1806$-$20
provides
direct evidence for a superfluid neutron core.  The highly non-thermal 
persistent X-ray emission of these sources must be powered
mainly by external currents; but a thermal seed could be provided
by internal heating.  (The AXP with the hardest X-ray spectrum 
and noisiest spindown,
1E 1048.1$-$5937, also has a relatively low blackbody luminosity,
less than $10^{34}$ ergs/s; \"Ozel et al. 2003.)
SGR 0526$-$66 has remained X-ray bright 
since its last observed outburst in 1983 (Kulkarni et al. 2003)
and has a high luminosity $\sim 7\times 10^{35}$ erg/s:  either
$B \gg 10^{15}$ G at its surface,
or most of its emission continues to be powered by external
currents (in spite of the relatively soft X-ray spectrum).
Detailed timing measurements of this source would provide valuable
diagnostics of its interior state.

\acknowledgments
We thank Lars Bildsten, Vicky Kaspi, Shri Kulkarni, Bennett Link, 
and Peter Woods for conversations.  PA is an AAPF NSF fellow,
AC is a Hubble Fellow, and CT acknowledges the support of
the NSERC of Canada.

\figurenum{1}
\begin{figure}
\plotone{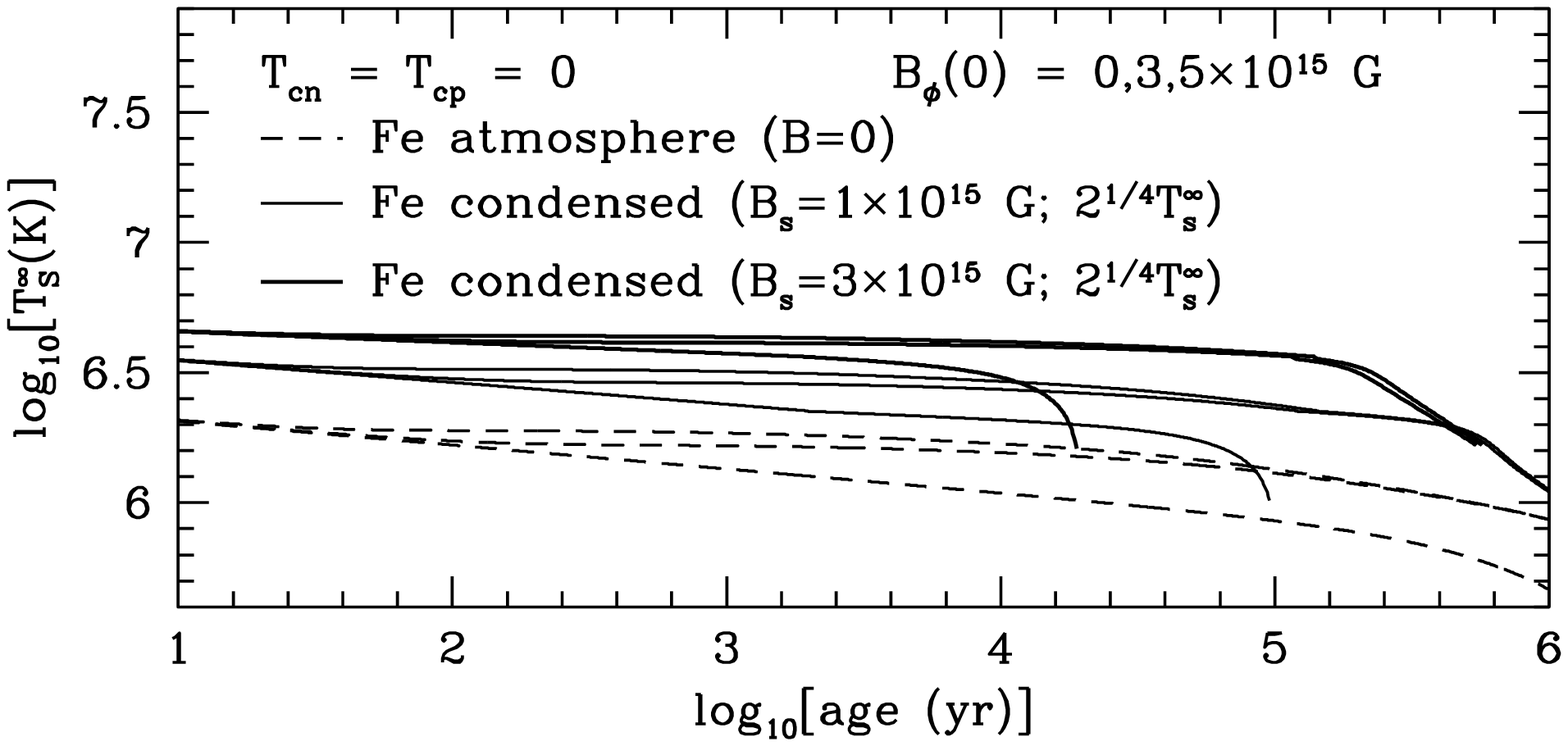}
\caption{Surface temperature (redshifted) as a function of age,
for a $1.35 M_\odot$ neutron star with normal neutrons and protons.
Upper two curves in each set correspond to stars with internal
heating mainly by ambipolar diffusion.)  $B_s$ is the surface field,
which is assumed to be constant in time, and 
$B_\phi$ the internal toroidal field (which may be stronger).
See text for the description 
of the envelope models. The factor of $2^{1/4}$ corrects for
surface emission dominated by the extraordinary polarization mode.}
\end{figure}

\clearpage

\figurenum{2}
\begin{figure}
\plotone{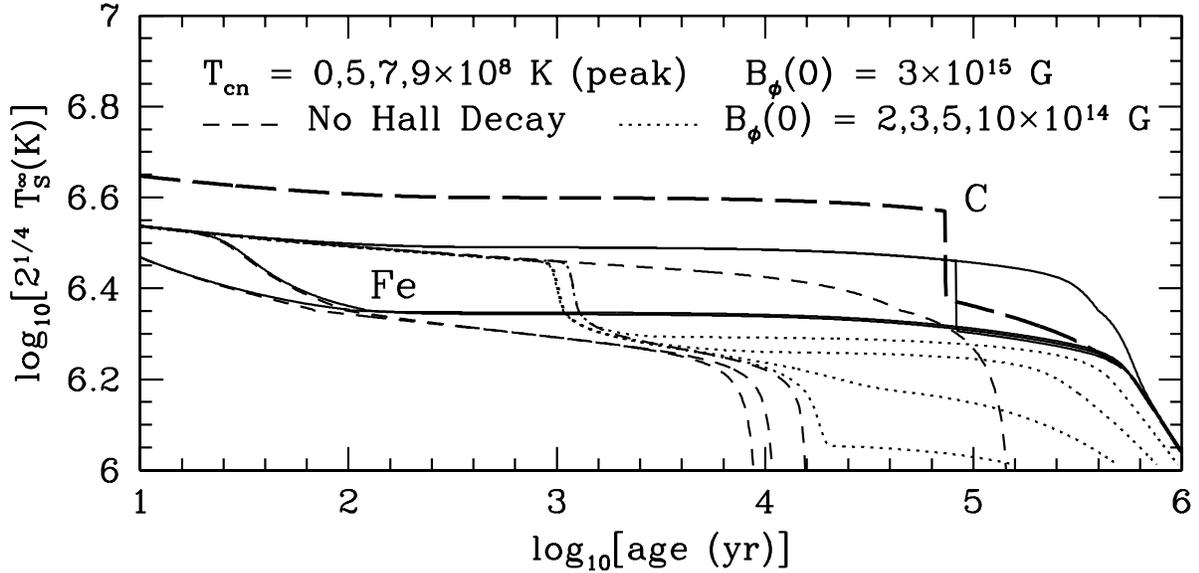}
\caption{Same but for a condensed iron envelope,
superconducting core protons ($T_{c,p} = 5\times 10^9$ K), 
and various peak pairing temperatures for the core neutrons.  
The distribution of critical temperature $T_{c,n}$ with density is 
taken to be $T_{c,n} = T_{c,n}({\rm peak})\exp[-(\rho_{14}-5)^2/\rho_{14}^2]$.
Smooth declines in X-ray flux at $t > 10^{4-5}$ yrs occur 
when surface cooling begins to dominate.  Sharper
drops signal a transition to core superfluidity.  Those occuring
at $t< 100$ yrs correspond to 
$T_{c,n}({\rm peak}) = 9,7\times 10^8$ K; and at $t \sim 10^5$ yrs
to $T_{c,n}({\rm peak}) = 5\times 10^8$ K.   Solid curves 
($B_s = 10^{15}$ G) include
the effects of Hall decay.  Dashed curves do not, and ordered from
left to right correspond to  $T_{c,n}({\rm peak}) = 9,7,5,0\times 10^8$ K. 
Notice that the superfluid transition is greatly delayed for 
$T_{c,n} = 5\times 10^8$ K.  Dotted curves show weaker magnetic fields,
$B_s = B_\phi(0) = 2-10\times 10^{14}$ G
and $T_{c,n} = 5\times 10^8$ K.  Finally, the top dashed curve shows
enhanced cooling through a condensed carbon envelope (density
$\leq 1\times 10^{10}$ g cm$^{-3}$) with $B_s = 
1\times 10^{15}$ G, $B_\phi(0) = 3\times 10^{15}$ G, 
and $T_{c,n}=5\times 10^8$ K.}
\end{figure}

\end{document}